# Evolution of Complexity in Out-of-Equilibrium Systems by Time-Resolved or Space-Resolved Synchrotron Radiation Techniques

**Gaetano Campi** [1,*] **and Antonio Bianconi** [1,2,3]

[1] *Institute of Crystallography, CNR, Via Salaria Km 29.300, Monterotondo, I-00015 Roma, Italy*
[2] *Rome International Center of Materials Science (RICMASS), Via dei Sabelli 119A, 00185 Roma, Italy; antonio.bianconi@ricmass.eu*
[3] *National Research Nuclear University MEPhI (Moscow Engineering Physics Institute), 115409 Moscow, Russia*

**\*** Correspondence: gaetano.campi@ic.cnr.it; Tel.: +39-06-9067-2624

**Abstract:** Out-of-equilibrium phenomena are attracting high interest in physics, materials science, chemistry and life sciences. In this state, the study of structural fluctuations at different length scales in time and space are necessary to achieve significant advances in the understanding of the structure-functionality relationship. The visualization of patterns arising from spatiotemporal fluctuations is nowadays possible thanks to new advances in X-ray instrumentation development that combine high-resolution both in space and in time. We present novel experimental approaches using high brilliance synchrotron radiation sources, fast detectors and focusing optics, joint with advanced data analysis based on automated statistical, mathematical and imaging processing tools. This approach has been used to investigate structural fluctuations in out-of-equilibrium systems in the novel field of inhomogeneous quantum complex matter at the crossing point of technology, physics and biology. In particular, we discuss how nanoscale complexity controls the emergence of high-temperature superconductivity (HTS), myelin functionality and formation of hybrid organic-inorganic supramolecular assembly. The emergent complex geometries, opening novel venues to quantum technology and to the development of quantum physics of living systems, are discussed.

**Keywords:** micro X-ray diffraction; synchrotron radiation; structural fluctuations; high-temperature superconductivity; superstructures; myelin; correlated disorder; supramolecular assembly

1. Introduction

The structural fluctuations at the nanoscale and mesoscale play a fundamental role in the functionality of complex materials [1–22]. The monitoring and understanding of the spatio-temporal dynamics require highly spatially resolved probes. Although standard momentum-space probes such as X-ray diffraction (XRD), angle-resolved photoelectron spectroscopy and neutron scattering are excellent for characterizing the "average" order and coherent excitations, they are highly ambiguous when different phases coexist and evolve in space and time. To overcome this limitation, local probes such as X-ray spectroscopy and pair distribution function analysis from high resolution X-ray and neutron diffraction have been used to probe local fluctuations and disorder in complex materials showing nanoscale phase separation [23–28].

Nowadays, thanks to the advanced features of the latest generation synchrotron sources and new X-ray optics [29–34], new techniques have been developed for this purpose. Scanning X-ray (sub)micro-diffraction (SµXRD) and spectroscopy, e.g., scanning micro X-ray absorption (SµXAS) and scanning micro X-ray fluorescence (SµXRF), constitute an optimal approach [35] to investigate structure fluctuations in several systems like perovskite materials such as $La_2CuO_{4+y}$ [36–43], $YBa_2Cu_3O_{6.5}$ [44–47], $K_{0.8}Fe_{1.6}Se_2$ [48–50]. Electronic fluctuations such as short-range Charge Density Waves (CDW) have been detected by resonant X-ray scattering [51–56] probing the average structure and by SµXRD [40,57–60] probing the spatial distribution of electronic nanometric patches. Scanning nano-probes with synchrotron radiation are now a hot topic in the investigation of structural fluctuations in biological systems [61–64].

In addition, fast detectors allow the in situ studies to monitor either physical or chemical modifications during data collection; for example, the few milliseconds readout time of the CMOS camera can be used for time-resolved measurements in different processes showing self-organization and pattern formation at the nanoscale [65–69]. Moreover, they allow us to prevent radiation damage on biological samples providing new key information [70–72].

In this work, we present some relevant results showing the relationship between new geometries developing at nano/mesoscale and the emerging macroscopic properties in complex and heterogeneous systems in different fields opening new fundaments in natural sciences [73–84].

2. Results

*2.1. High $T_c$ Superconductivity: Scale-Free Oxygen Distribution in $La_2CuO_{4+y}$*

High-temperature superconducting perovskites are heterostructures at the atomic limit composed of parallel planes of active layers (copper oxide in the case of cuprates) sandwiched between rock-salt layers. The copper atoms lie on the plane where the charge is carried by "holes" compensated by dopants in the spacer layers. It is now accepted that the interplay of defects, charge, spin and orbitals play a fundamental role in copper-oxide-based superconductors, belonging to the wide family of *quantum complex materials* [1–22]. Early X-ray spectroscopy measurements XANES and EXAFS of copper oxide compounds revealed nanoscale local lattice fluctuations giving tilts and bond disproportionated $CuO_4$ plaquettes [25] with the $\{3d^9\underline{L}\} - 2d^{10}\underline{L}^2\}$ (where $\underline{L} = O(2p^5)$ and $\underline{L}^2 = O(2p^4)$) many body electronic configurations, associated with doped formal $Cu^{3+}$ impurities, which are formed in the charge transfer correlation gap between the $\{3d^9\}$ and $\{3d^{10}\underline{L}\}$ in undoped $Cu^{2+}$ oxides.

Among high-temperature superconductors (HTS) the most simple compound is $La_2CuO_{4+y}$ where y oxygen interstitial ions (O-i), are inserted in the rocksalt $[La_2O_{2+y}]$ intercalated by $[CuO_2]$ layers [36–43]. Due to the mobility of oxygen interstitials, O-i, this compound exhibits a rich phase diagram. The multiscale structural conformations due to the dynamic distribution of mobile O-i have been studied by X-ray synchrotron diffraction. Thanks to the high photon flux, we have measured the weak "diffuse scattering" in a single-crystal of $La_2CuO_{4+y}$ with orthorhombic Fmmm symmetry. Beyond the Bragg peaks, we have detected satellite peaks, associated with super-cells due to O-i dopants arrangement [36–39]. These O-i ordered domains show an intriguing inhomogeneous spatial distribution (see the maps in Figure 1a,b) as seen by space resolved SμXRD measurements [40–42]. We have characterized this inhomogeneity by the (i) probability density function and (ii) spatial correlation function (see Methods).

Figure 1c,d (green regions) shows the probability distribution of the O-i domains in two different samples: the first one, named $T_c$ = 40 K, has a superconducting $T_c$ of 40 K while the second one presents a mixture of two superconducting phases with $T_c$ of 16 K and 32 K, here named $T_c$ = 16 K. In both cases we find a fat tail above the average intensity, <I>, following a clear power-law distribution with exponential cut-off, $x_0$, given by

$$P(x) \sim x^{-\alpha}\exp(-x/x_0). \qquad (1)$$

The power-law exponent, $\alpha$, is always indistinguishable from 2.6 while the cut-off, $x_0$, is less than 10 for the $T_c$ = 16 K materials and greater than 10 for the high-$T_c$ = 40 K materials.

The spatial correlation function, G(r) (orange regions of Figure 1c,d), where $r = |Ri - Rj|$ is the distance between x–y positions on the sample and also follows a power law with a cut-off, as expected near a critical point:

$$G(r) \sim r^{-\xi}\exp(-r/\xi) \qquad (2)$$

with ξ = 0.3 ± 0.1 for all samples. The correlation length, ξ, increases with increasing <I>, varying between 50 and 250 mm for the $T_c$ = 16 K samples and taking the respective values 400 ± 30 mm and 350 ± 30 mm for the $T_c$ = 40 K samples. The results for both the intensity distribution and the two-point correlation function show that the unexpected fractal nature of O-i ordering, associated with the measured power laws, is robust, approaching a pure scale-free distribution in the sample with higher critical temperature [34].

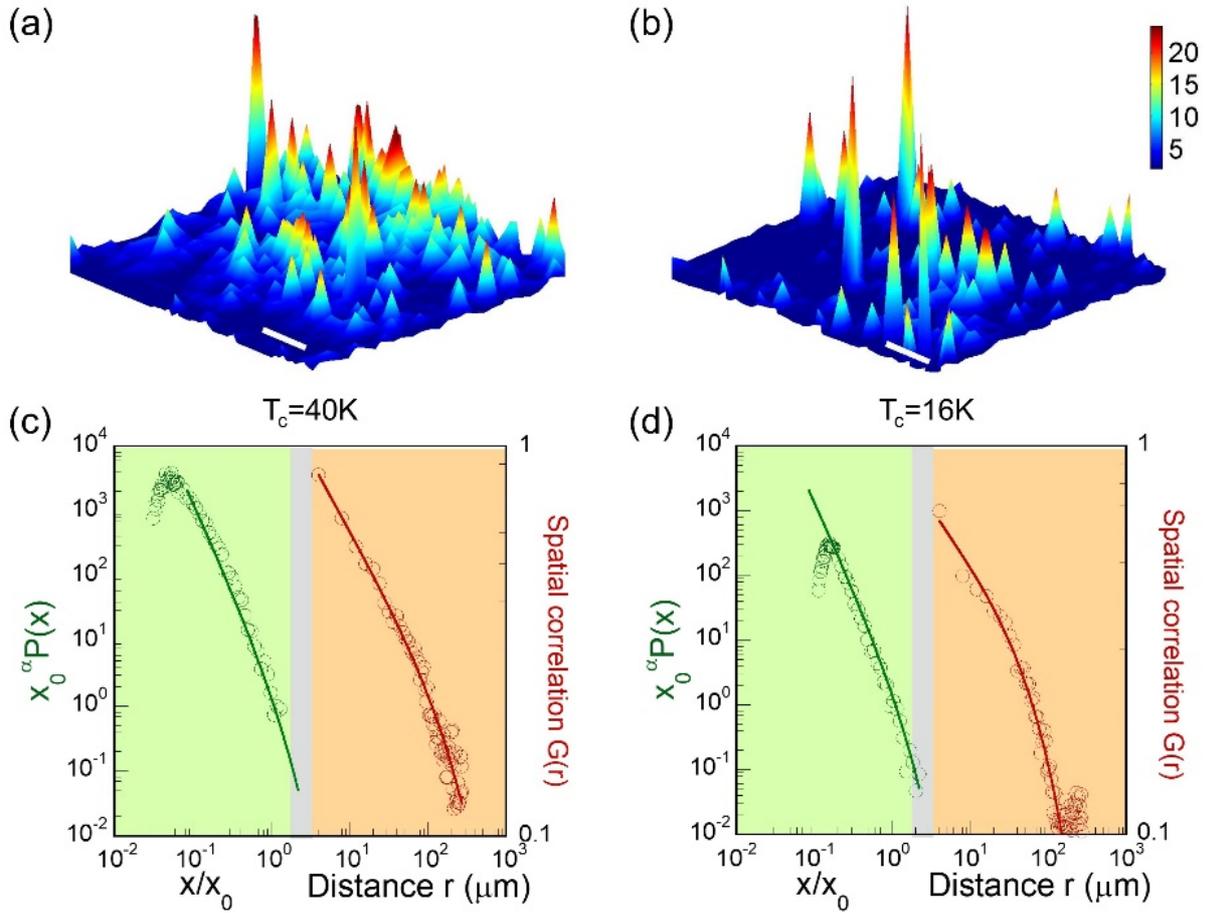

**Figure 1.** The position dependence of the superstructure intensity for two typical samples with
$T_c$ = 40 K (**a**) and $T_c$ = 16 K (**b**) phases in a 100 × 100 μm² area. The white bar corresponds to 20 μm. The probability distribution, P($x$), and spatial correlation function, G($r$) of Q2 for the same two typical samples, are presented in panel (c) and (d), respectively.

*2.2. High $T_c$ Superconductivity: Scale Free Oxygen Distribution and Charge-Density-Waves in HgBa₂CuO₄₊ᵧ*

Alongside the oxygen ordering, the electronic inhomogeneity, namely charge density waves (CDW), has been measured in the HgBa₂CuO₄₊ᵧ [58]. We have been able to measure the diffuse scattering associated with both CDW and oxygen interstitial arrangement in the lattice.

In Figure 2a we show the maps of the integrated intensity of (left panel) CDW peak and (right panel) oxygen O-i diffuse scattering. Both maps clearly show spatial inhomogeneity described by the PDFs assuming a power-law behavior, also in this case, as shown in Figure 3b. However, the critical exponent in the two distributions assumes different values: 1.8 ± 0.1 and 2.2 ± 0.1 in the O-i and CDW intensity, respectively. The other interesting feature of this spatial inhomogeneity is the negative spatial correlation between O-i and CDW. This is well depicted in the 'difference map' between CDW peaks and O-i diffuse streaks shown in Figure 3c..

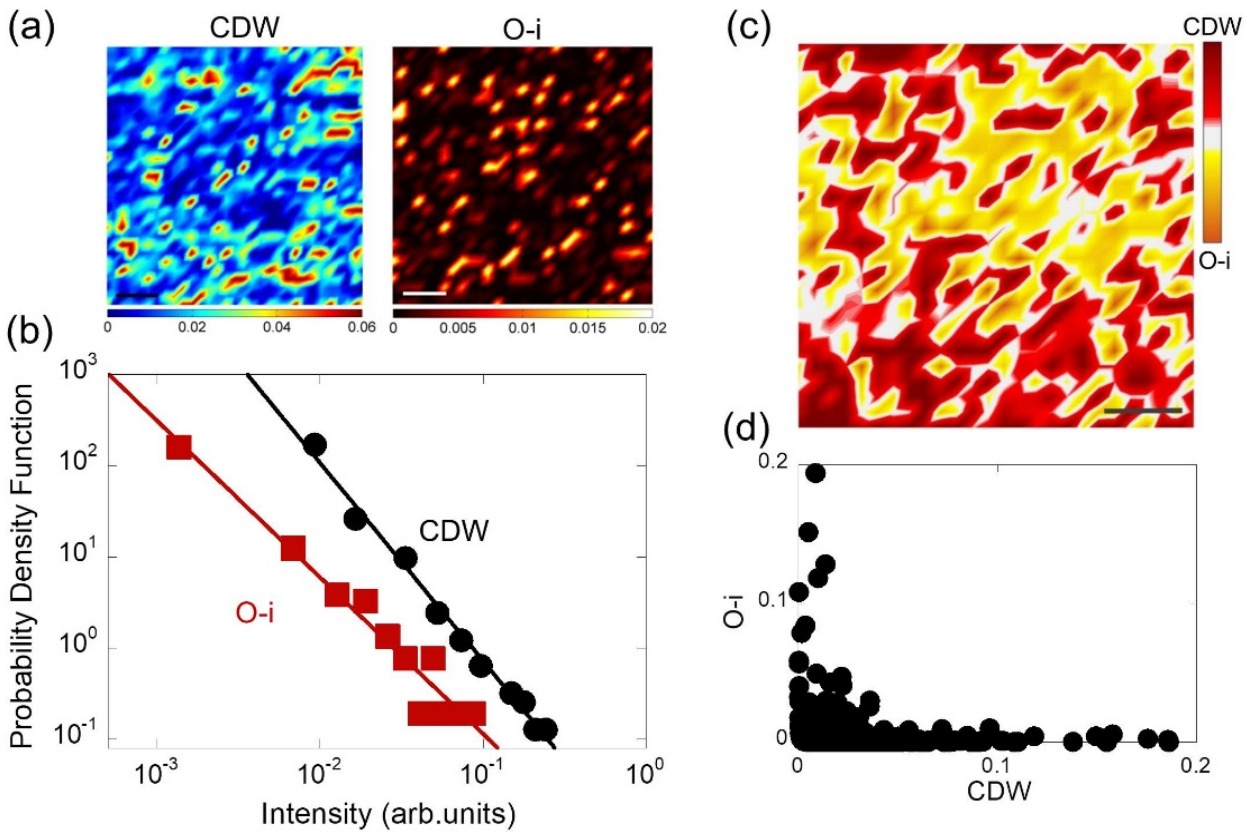

**Figure 2.** (**a**) Color plot of the difference map between the Charge Density Waves (CDW)-peak and O-i-streak intensity. The positive red values indicate the CDW-rich regions and the negative yellow values correspond to O-i rich regions. The scale bar corresponds to 5 μm. (**b**) Probability density function calculated from the Oi-streaks and CDW intensity map. (**c**) Difference map between O-i and CDW highlighting the different O-i rich zones (red) and CDW puddles (yellow). The scale bar corresponds to 5 μm. (**d**) Scatter plot of O-i versus CDW intensity demonstrating the negative correlation between CDW-puddle and O-i.

The poor CDW regions on the $CuO_2$ basal plane correspond to O-i rich regions on the $HgO_y$ layers. The CDW puddles and O-i rich regions give rise to the positive (red) and negative (yellow) zones, respectively. Here we can highlight the interface, given by the white region, between the CDW puddles and O-i quenched disorder where interference between different pathways in a non-euclidean, hyperbolic geometry [77,78] can help to raise the critical temperature. This spatial negative correlation is also evident from the scatter plot of O-i intensity versus CDW intensity (Figure 3d). These results open new venues in the double-handed issue pertaining both to technology, relative to controlled fabrication of superconducting devices and to basic physics dealing with the formation and evolution of quantum coherence in relation to complex geometries at the nanoscal

*2.3. Biology: Ultrastructural Fluctuations in Biological Systems*

Myelin can be considered a simple example of a biological ultrastructure. The scheme of the multilamellar ultrastructure of myelin is shown in Figure 3a. It is made of the stacking of (i) cytoplasmatic (cyt), (ii) lipidic (lpg), (iii) extracellular (ext) and iv) another lipidic (lpg) layer [61,62]. The individual thickness of each layer in a 1 μm² spot area, named $d_\lambda$, $d_{cyt}$, $d_{lpg}$, $d_{ext}$, has been extracted from electron density profiles computed by Fourier analysis of the diffraction patterns, as described in details in Campi et al. [61].
The structural fluctuations of myelin stacks and their spatial correlations have been quantified by introducing the appropriate conformational parameter, ξ, given by the ratio between hydrophilic and hydrophobic layers:

$$\xi = (d_{ext} + d_{cyt})/2d_{lpg} \quad (3)$$

This parameter characterizes univocally the state of myelin. Typical maps of ξ measured on a 100 × 100 μm² ROI of a nerve in the functional living phase and in the aged resting phase are shown in Figure 3b,c, respectively. The red spots represent areas where hydrophilic layers are larger, while in the blue spots they become smaller and the thickness of the hydrophobic layer increases. The PDF of ξ in the living phase shows a skewed line shape modelled by using Levy stable distributions (see Figure 3d). The Levy stable distribution provide a general statistical description of complex signals deviating from normal behavior and in recent years have found increasing interest in several applications in diverse fields [85,86]. This class of probability distributions are generally represented by a characteristic function defined by four parameters: stability index $\alpha$, skewness parameter $\beta$, scale parameter $\gamma$ taking into account the width of the distribution, and location parameter $\delta$ with varying ranges of $0 < \alpha \leq 2$, $-1 \leq \beta \leq 1$, $\gamma > 0$ and $\delta$ real. Here we have used basic functions in the numerical evaluation of these parameters and goodness of data fitting as described by Liang and Chen [87]. We stress the fact that the closed-form expressions of density and distribution functions of Levy stable distributions are not available except for few particular cases such as the well-known normal distribution where the stability parameter $\alpha$ is 2, as occurs in the resting phase in the aged sample. In the functional living state, the ξ distribution function follows a Levy fitting curve, indicated by the continuous line in Figure 3d, with a stability index of 1.78 (<2), a location of 0.8242, a skewness of 1 and a scale parameter $\gamma$ of 0.0138.

*2.4. Hybrid Organic-Inorganic Supramolecular Assembly*

In situ time-resolved small angle X-ray scattering (SAXS) has been used to monitor hybrid particle diffusion and aggregation at the nanoscale. We consider here the case of silver nanoparticle synthesized in aqueous solution with polynaphtalene sulphonate polymer as stabilizer [65–67]. We have monitored the particles formation by adding ascorbic acid as a reducing agent at a fixed slow rate, in order to control the Ag+ ions reduction, achieving the formation of the $Ag^0$ crystalline colloidal dispersion.

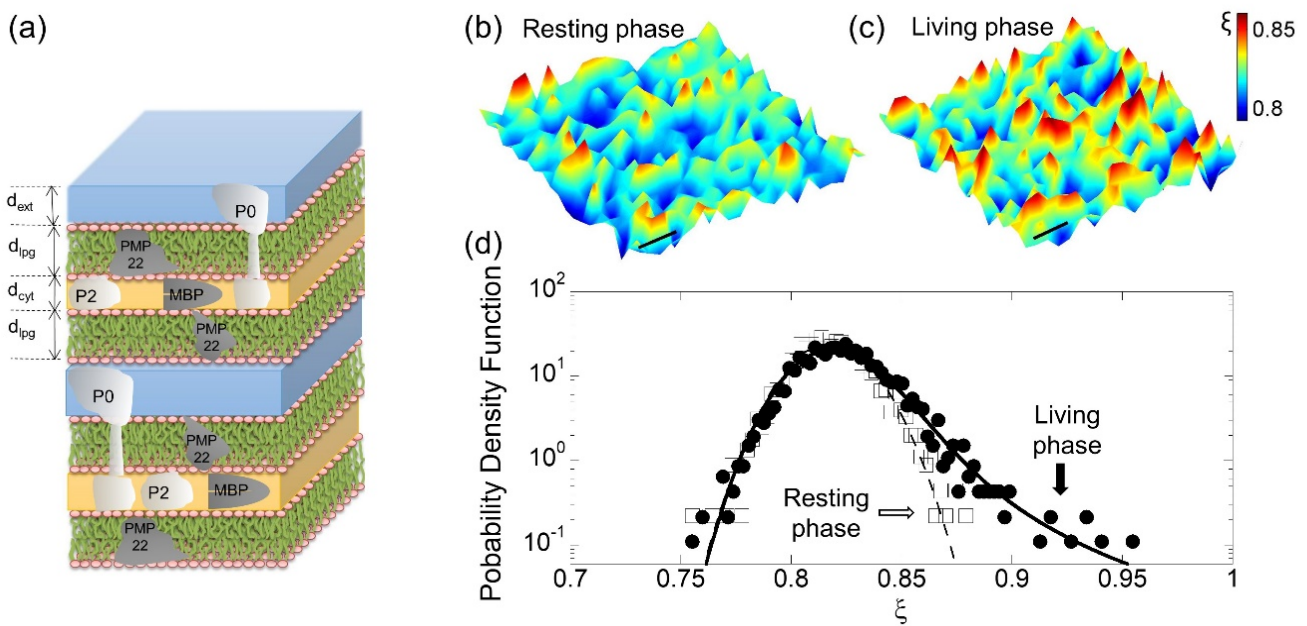

**Figure 3.** (**a**) Pictorial view of the protein depleted membrane layers made of polar lipid groups, lpg, with thickness $d_{lpg}$, intercalated by two hydrophilic layers: the Schwann cell cytoplasm, cyt, and the extracellular apposition, ext, with thickness $d_{cyt}$ and $d_{ext}$, respectively. The specific myelin sheath protein PMP22, P0, P2 and MBP are schematized. Map of the conformational parameter, ξ, in a selected central zone of the (**b**) unfresh and (**c**) fresh nerve (400 × 125 μm²). (**d**) The probability density function of ξ in the unfresh (open squares) and in the fresh (full circles) sample in semi-log plot. The Levy Probability Density Function curve found in the fresh sample (continuous line) is reported. We notice the loosing of the fat tail in the distribution assuming a Gaussian profile (dashed line) in the unfresh sample.

We have used SAXS at *T* = 40 °C in the 0.07 < *q* < 1.7 nm$^{-1}$ range using a CCD detector with a time resolution of 0.9 s. SAXS normalized and water-subtracted intensity versus *q* at different time intervals are shown in Figure 4a. Before the addition of ascorbic acid, the scattered intensity, due to the Ag salt + polymer background in the aqueous solution, shows power law profiles (see the SAXS pattern collected at *t* = 22.5 s in Figure 3a). As the first micro drop of reduction agent falls in the solution mixture, the scattered intensity starts to gradually increase in the low *q* region indicating particle nucleation and growth. In the following minutes we detect (i) the development of interference oscillations whose maxima and minima progressively shift to smaller *q* and (ii) a change in the profile slopes at larger *q* that increase as a function of the time (Figure 4a).

These essential features of nucleation and growth mechanism are captured by assuming a physical model with two components constituted by interacting spherules forming in an aqueous solution of Ag salt and polymer [66–69] template. The first component, $I_B(q) \sim q^{P_E}$ is ascribed to the polymeric template in the solution; the second one describes highly dispersed particles interacting via hard-sphere potential modelled by Percus-Yevick structure factor $S(q, R_{HS}, \eta)$ where $R_{HS}$ and $\eta$ represent the correlation length and the volume fraction of interacting particles, respectively [66,67].

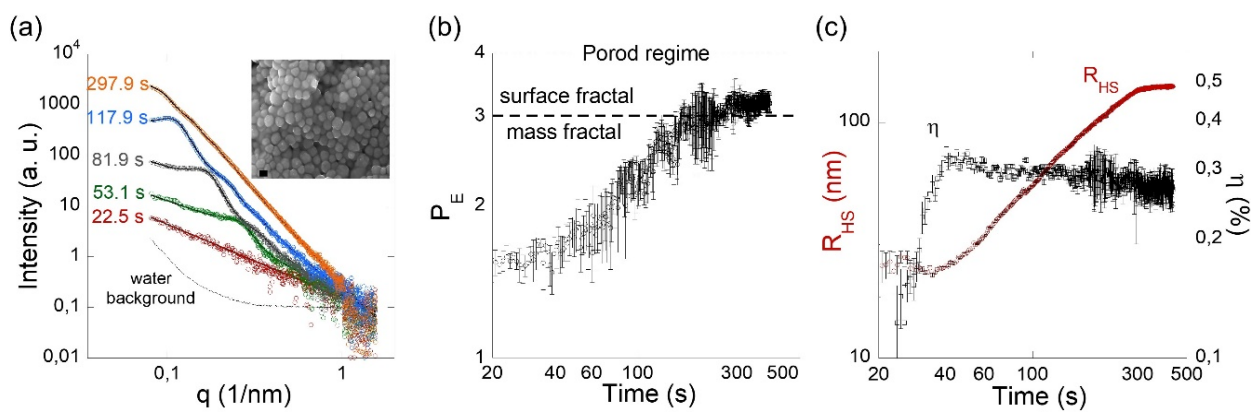

**Figure 4.** (**a**) Small Angle X Ray Scattering normalized profiles (open circles) collected at the time intervals indicated, at *T* = 40 °C; solid lines show the best-fitted curves calculated using the two components model described in the text. In the inset we show an image of formed nanoparticles obtained by Transmission Electron Microscopy (**b**) The fractal background dimension is given by the exponent, $P_E$; confidence intervals from optimization procedure are indicated with error bars. (**c**) Time evolution of correlation length $R_{HS}$ and volume fraction $\eta$.

Time evolution of both power law exponent, $P_E$, associated with polymer-Ag matrix fluctuations, and structure factor taking into account the inter-particle interactions, are shown in Figure 4b,c respectively. We can observe how fractal dimension changes leading to a mass-surface fractal transition; at the same time the correlation length increases up to very large distances of about 120 nm, comparable with the final nanoparticle shown by the Transmission Electron Microscopy (TEM) image in the inset of Figure 4a. This procedure gives rise to different geometries at the nanoscale, just by changing the temperature [67]. This wet easy approach has been used for different hybrid organic-inorganic mixtures, where interactions developing between structural units of supramolecular assembly give rise to complex geometries down to few nanometers, with intriguing functionality for applications in different fields [68,69].

## 3. Discussion

A wide variety of systems in nature work out of thermodynamic equilibrium where spatial and time fluctuations give rise to complex patterns showing correlated disorder at nanoscale and mesoscale. The visualization and understanding of this disorder represent a challenge in experimental as well as theoretical science. Although self-organizing criticality and fractal geometry has been the preferred description for these phenomena, new, more complex and unexpected insights enrich this field of research as the emergent role of hyperbolic geometry determining the functionality of complex networks

In our approach, we first visualize the spatial and time fluctuations at nano and mesoscale using time and space resolved synchrotron radiation techniques; afterwards, we quantify these fluctuations using advanced statistical physics.

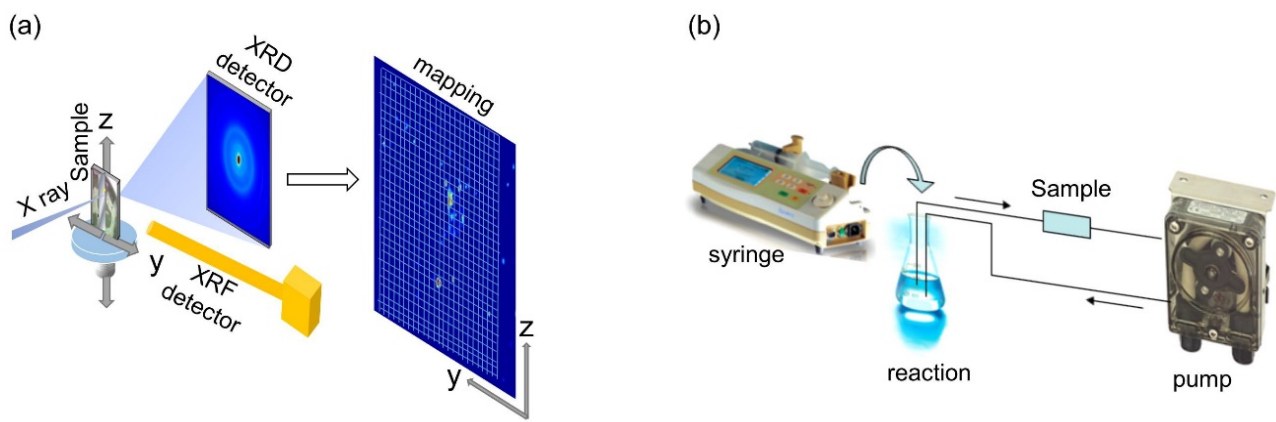

**Figure 5.** Typical X-ray diffraction and spectroscopy apparatus for (**a**) space and (**b**) time-resolved measurements.

What we find is that nanoscale and mesoscale fluctuations follow a normal behavior in the equilibrium state. In the out of equilibrium status, where all the biological matter works, these fluctuations deviate from normal behavior assuming fat-tailed distributions, relate to the emergence of a correlation degree in the disorder. We have found that these fluctuations show a power law in the HTS and a Levy behavior in the myelin. This can be particularly important in the biomedical field where the approaching of morphological fluctuations to a normal behavior can indicate a pathological condition. This could be used as new preclinical early stage detection of degenerative processes and disease.

## 4. Materials and Methods

*4.1. Experimental Set Up for Space-Time Resolved Measurements*

A typical space resolved experimental setup with synchrotron radiation includes an in-vacuum undulator as the primary source, mirrors for beam focusing, monochromators for energy selection and focusing optics such as compound refractive lenses (CRLs) [29,30], Kirkpatrick Baez (KB) mirrors [31], crossed Fresnel zone plates [32] and waveguides [33,34]. The different optics provide different beam spot sizes ranging from 50 nm to few microns. In Figure 5a is illustrated the scheme of a typical experimental apparatus for space-resolved X-ray scattering in transmission geometry.

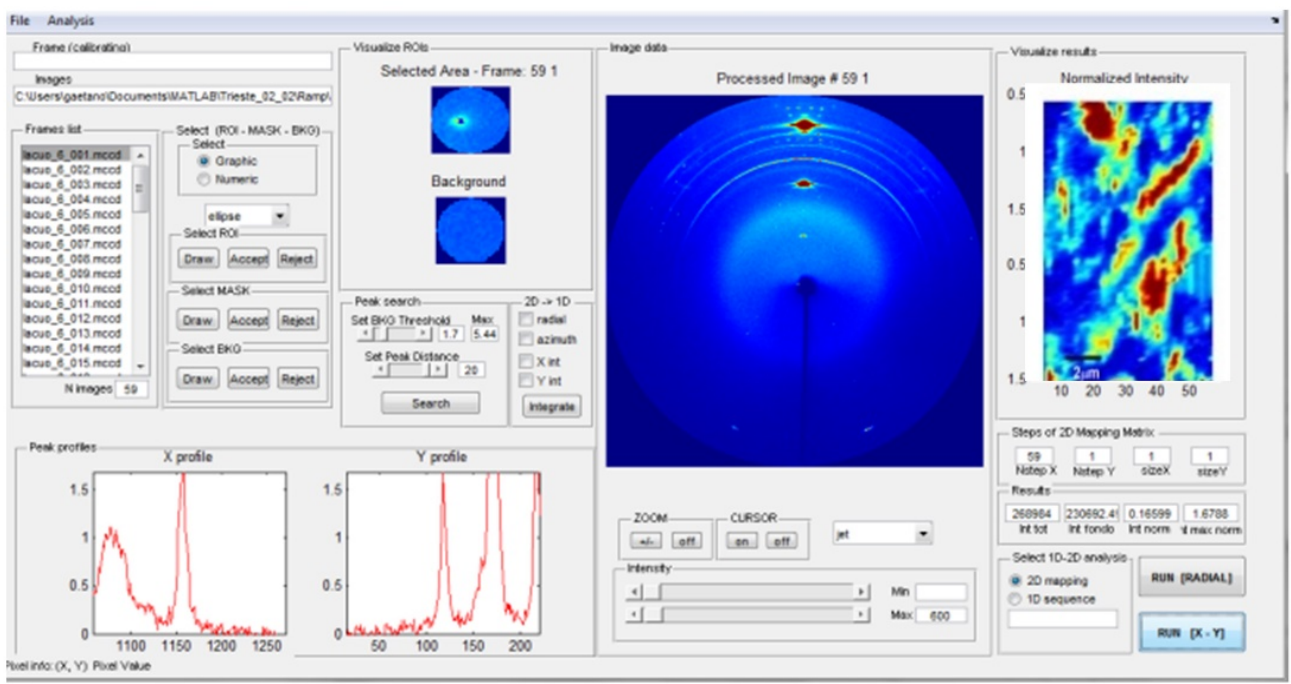

**Figure 6.** Main panels of the software *MapX* for data analysis. The program allows the following operations: to download a big number of measured images; to select a diffraction feature such as a peak or some diffuse scattering; to integrate and build X-ray profiles that can be fitted by appropriate line-shapes. Finally, one cane builds spatial maps of each parameter extracted from the fit.

The scanning micro X-ray diffraction measurements of HTS and myelin of frog's sciatic nerve were performed on the ID13 beamline of the European Synchrotron Radiation Facility, ESRF, France. The experimental methods were carried out in "accordance" with the approved guidelines. The source of the synchrotron radiation beam is an 18 mm period in a vacuum undulator. The beam is first monochromatized by a liquid nitrogen cooled Si-111 double monochromator (DMC) and then is focused by the diffractive lens.

These optics produce an energy X-ray beam of $\lambda$=12.6 KeV on a 1 × 1 µm² spot. The sample holder allows both horizontal (y) and vertical (z) translation stages with 0.1 µm repeatability.

A Fast Readout Low Noise (FReLoN) camera (1024 × 1024 pixels of 100 × 100 µm²) was used for data collection. In the case of myelin, we chose an exposure time of 300 ms for minimizing the radiation damage and for keeping a good sensitivity at the same time. 2-D diffraction patterns with the expected arc-rings corresponding to the Bragg diffraction orders h = 2, 3, 4, 5 were measured [61,62]. The electron density profiles were extracted by diffraction intensities using Fourier analysis [61].

The $La_2CuO_{4+y}$ (LCO) single crystal with *y* = 0.1 has orthorhombic *Fmmm* space group symmetry with lattice parameters a = (5.386 ± 0.004) Å, b = (5.345 ± 0.008) Å, c = (13.205 ± 0.031) Å at room temperature. The $HgBa_2CuO_{4+y}$ (Hg1201) single crystal with y = 0.12 has a sharp superconducting transition at $T_c$ = 95 K. The crystal structure has tetragonal P4/mmm space group symmetry with lattice parameters a = b = 0.387480(5) nm and c = 0.95078(2) nm at *T* = 100 K. Diffraction measurements on single crystals of LCO and Hg1201, were performed on the ID13 beamline at ESRF.

The formation dynamics of chemical hybrid dispersions was monitored by in situ time-resolved measurements. Nowadays, the combination of microfluidics and SAXS allows us to investigate dynamic processes down to millisecond time resolution. In this case study, the batch reactor apparatus consisted of a glass flask, a remote-controlled syringe that allowed us to add 4 mL of reducing ascorbic acid solution at the fixed rate of 0.5 mL min$^{-1}$, and a peristaltic pump that continuously flows the solution mixture in a 1.5 quartz capillary through a closed circuit (see Figure 5b).

The pumping rate was set to 20 mL min$^{-1}$ in order to change all the tubing (1 m × 2 mm) volume in less than 10 s, avoiding particle deposition on the walls. The exposure time for collecting each frame was 0.9 s.

*4.2. Program MapX to Analyze X-ray Diffraction and Spectroscopy Big-Data Sets*

If we consider a sample area of 100 × 100 µm, measured with a step of 100 nm in both y and z directions, we collect 1,000,000 patterns. At the same time, if we measure a slow chemical reaction taking place in 10 min, with a time resolution of 10 ms, we need to collect 60,000 frames.

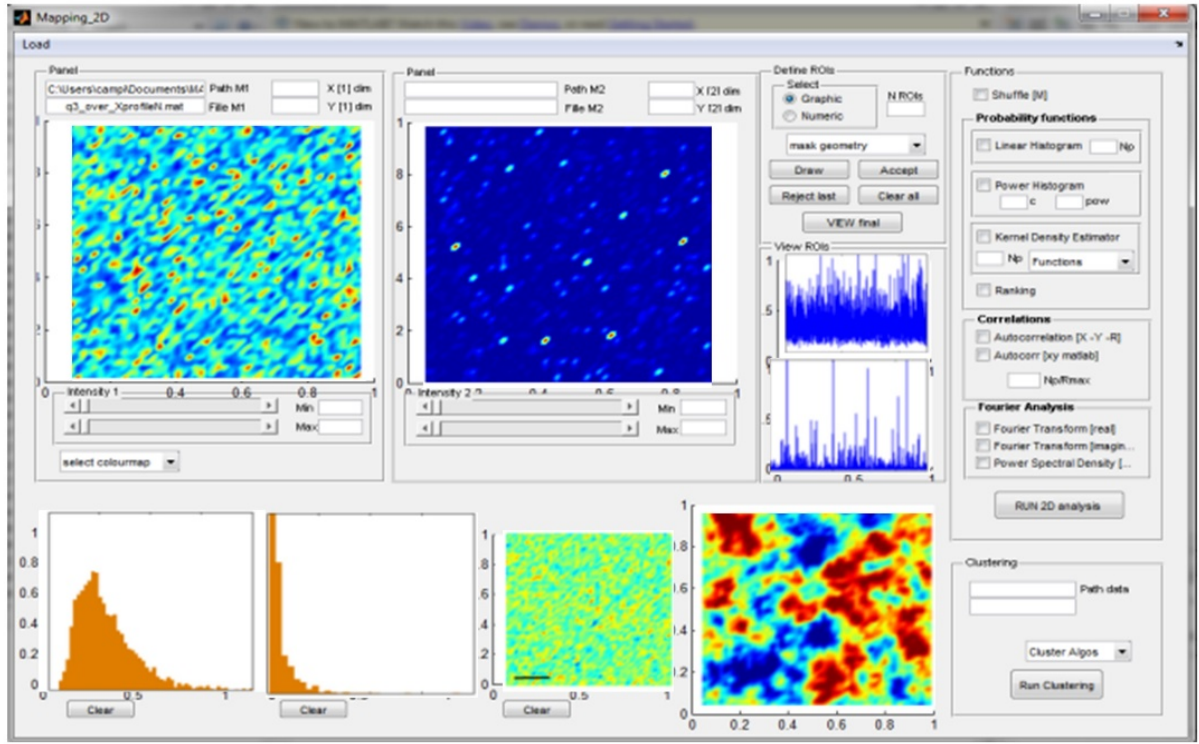

**Figure 7.** Panels of the software *MapX* for spatial statistical data analysis. The geometry at nanoscale and mesoscale is characterized by calculating several physical spatial statistical quantities, such as the probability density function, spatial correlations, clustering degree, tessellation, percolation.

Thus, the large quantity of collected data gives rise to new problems dealing with a big-data set to store and to analyze. For this aim, a dedicated software, *MapX*, has been developed and written in Matlab. In Figure 6 we show the main panel of this program for X-ray diffraction and spectroscopy big data sets analysis. It allows us to select specific features of each single diffraction pattern and to analyze and fit them to extract several parameters of interest, e.g., peak width, intensity or position. Once extracted, these parameters are used to build spatial maps.

The second panel, shown in Figure 7, allows us to perform spatial statistical analysis of the obtained maps. The geometry at the nanoscale and mesoscale is characterized through several physical statistical quantities, such as the probability density function, spatial correlation function, clustering degree, tessellation, percolation. Different maps and map ROIs can be selected, calculating also cross-correlations.

## 5. Conclusions

We have discussed the key role of nanoscale and mesoscale geometry arising from structural fluctuations in some simple model systems working in out-of-equilibrium functional conditions: High-temperature Superconducting $La_2CuO_{4+y}$ and $HgBa_2CuO_{4+y}$ crystals, myelin ultrastructure and hybrid nanostructures. In high-temperature superconductors, the quantum coherence is related to universal nanoscale fluctuations [36–60,88–92], where scale-free nanoscale phase separation create filamentary superconductivity in hyperbolic space [77,78]. In the myelin the Levy nanoscale ultrastructural fluctuations take place thanks to the correlated dynamics of cytosolic and lipid biological membranes, ensuring the physiological for the electrical transmission in nerves. In hybrid organic-inorganic nanostructures, we have visualized the evolution of supramolecular assembly associated with the mass-surface fractal transition. These studies have been possible thanks to the new advanced features of the latest generation synchrotron sources and fast acquisition detectors. The different statistical physics found in these out-of-equilibrium systems tells us that the visualization of fluctuations can give us fundamental information about functionality in new advanced materials.

**Author Contributions:** G.C. and A.B. conceived and designed the work.

**Funding:**  This research received no external funding

**Acknowledgments:** The authors thank Luisa Barba, Heinz Amenitsch and XRD1 and SAXS beamlines staff at ELETTRA, Trieste, Italy. The authors thankManfred Burghammer and ID13 beamline staff at ESRF, Grenoble, France. We are grateful to Alessandro Ricci, Nicola Poccia, Michela Fratini, Michael Di Gioacchino and Lorenza Suber for long term scientific collaboration.

**Conflicts of Interest:** The authors declare no conflict of interest.